\documentclass[preprint]{emulateapj}
\usepackage{hyperref}
\usepackage[usenames,dvipsnames]{color}
\hypersetup{colorlinks,citecolor=Blue,linkcolor=Red,urlcolor=Blue}
\usepackage[titletoc]{appendix}
\usepackage{amsmath}

\newcommand{\be}{\begin{equation}}
\newcommand{\ee}{\end{equation}}
\def\lta{\,\raise 0.3 ex\hbox{$ < $}\kern -0.75 em
 \lower 0.7 ex\hbox{$\sim$}\,}
\def\gta{\,\raise 0.3 ex\hbox{$ > $}\kern -0.75 em
 \lower 0.7 ex\hbox{$\sim$}\,} 

\begin{document} 

\title{Spin-orbit misalignment as a driver of the \textit{Kepler} Dichotomy} 

\author{Christopher Spalding$^1$ and Konstantin Batygin$^1$} 
\affil{$^1$Division of Geological and Planetary Sciences\\
California Institute of Technology, Pasadena, CA 91125} 
\affil{$\,$}

\begin{abstract}
During its 5 year mission, the \textit{Kepler} spacecraft has uncovered a diverse population of planetary systems with orbital configurations ranging from single-transiting planets to systems of multiple planets co-transiting the parent star. By comparing the relative occurrences of multiple to single-transiting systems, recent analyses have revealed a significant over-abundance of singles. Dubbed the ``\textit{Kepler} Dichotomy," this feature has been interpreted as evidence for two separate populations of planetary systems: one where all orbits are confined to a single plane, and a second where the constituent planetary orbits possess significant mutual inclinations, allowing only a single member to be observed in transit at a given epoch. In this work, we demonstrate that stellar obliquity, excited within the disk-hosting stage, can explain this dichotomy. Young stars rotate rapidly, generating a significant quadrupole moment which torques the planetary orbits, with inner planets influenced more strongly. Given nominal parameters, this torque is sufficiently strong to excite significant mutual inclinations between planets, enhancing the number of single-transiting planets, sometimes through a dynamical instability. Furthermore, as hot stars appear to possess systematically higher obliquities, we predict that single-transiting systems should be relatively more prevalent around more massive stars. We analyze the \textit{Kepler} data and confirm this signal to be present.
\end{abstract}

\section{Introduction} 

In our Solar System, the orbits of all 8 confirmed planets are confined to the same plane with an RMS inclination of $\sim$1-2$^\circ$, inspiring the notion that planets arise from protoplanetary disks \citep{Kant1755,Laplace1796}. By inference, one would expect extrasolar planetary systems to form with a similarly coplanar architecture. However, it is unknown whether such low mutual inclinations typically persist over billion-year timescales. Planetary systems are subject to many mechanisms capable of perturbing coplanar orbits out of alignment, including secular chaos \citep{Laskar1996,Lithwick2012}, planet-planet scattering \citep{Ford2008,Beauge2012} and Kozai interactions \citep{Naoz2011}.

Despite numerous attempts, mutual inclinations between planets are notoriously difficult to measure directly \citep{Winn2015}. In light of this, investigations have turned to indirect methods. For example, by comparing the transit durations of co-transiting planets, \citet{Fabrycky2014} inferred generally low mutual inclinations ($\sim1.0-2.2^\circ$) within closely-packed \textit{Kepler} systems. Additionally, within a subset of systems (e.g., 47 Uma and 55 Cnc) stability arguments have been used to limit mutual inclinations to $\lesssim40^\circ$ \citep{Laughlin2002,Veras2004,Nelson2014}. On the other hand, \citet{Dawson2014a} have presented indirect evidence of unseen, inclined companions based upon peculiar apsidal alignments within known planetary orbits. Obtaining a better handle on the distribution of planetary orbital inclinations would lend vital clues to planet formation and evolution.

 Recent work has attempted to place better constraints upon planet-planet inclinations at a population level, by comparing the number of single to multi-transiting systems within the \textit{Kepler} dataset \citep{Johansen2012,Ballard2016}. Owing to the nature of the transit technique, an intrinsically multiple planet system might be observed as a single if the planetary orbits are mutually inclined. An emerging picture is that although a distribution of small $\sim 5^\circ$ mutual inclinations can explain the relative numbers of double and triple-transiting systems, a striking feature of the planetary census is a significant over-abundance of single-transiting systems. Furthermore, the singles generally possess larger radii (more with $R_{\textrm{p}}\gtrsim 4$ Earth radii), drawing further contrast.

The problem outlined above has been dubbed the ``\textit{Kepler} Dichotomy," and is interpreted as representing at least two separate populations; one with low mutual inclinations and another with large mutual inclinations that are observed as singles. The physical origin of this dichotomy remains unresolved \citep{Morton2014,Becker2016}. To this end, \citet{Johansen2012} proposed the explanation that planetary systems with higher masses undergo dynamical instability, leaving a separate population of larger, mutually inclined planets, detected as single transits. While qualitatively attractive, this model has two primary shortcomings. First, it cannot explain the excess of smaller single-transiting planets. Second, unreasonably high-mass planets are needed to induce instability within the required $\sim$Gyr timescales. Accordingly, the dichotomy's full explanation requires a mechanism applicable to a more general planetary mass range. In this paper we propose such a mechanism - the torque arising from the quadrupole moment of a young, inclined star.

 The past decade has seen a flurry of measurements of the obliquities, or spin-orbit misalignments, of planet-hosting stars \citep{Winn2010,Albrecht2012,Huber2013,Morton2014,Mazeh2015,Li2016}. A trend has emerged whereby hot stars ($T_{\textrm{eff}}\gtrsim 6200$\,K) hosting hot Jupiters possess obliquities ranging from 0$^\circ$ to 180$^\circ$, as opposed to their more modestly inclined, cooler (lower-mass) counterparts. Further investigation has revealed a similar trend among stars hosting lower-mass and multiple-transiting planets \citep{Huber2013,Mazeh2015}. Most relevant to the \textit{Kepler} Dichotomy, \citet{Morton2014} concluded at 95\% confidence that single-transiting systems possess enhanced spin-orbit misalignment compared to multi-transiting systems. 

Precisely when these spin-orbit misalignments arose in each system's evolution is still debated \citep{Albrecht2012,Lai2012,Storch2014,Spalding2015}. However, the presence of stellar obliquities within currently coplanar, multi-planet systems hints at an origin during the disk-hosting stage \citep{Huber2013,Mazeh2015}. Indeed, many studies have demonstrated viable mechanisms for the production of disk-star misalignments, including turbulence within the protostellar core \citep{Bate2010,Spalding2014b,Fielding2015} and torques arising from stellar companions \citep{Batygin2012,Batygin2013,Spalding2014a,Lai2014,Spalding2015}. Furthermore, \citet{Spalding2015} proposed that differences in magnetospheric topology between high and low-mass T Tauri stars \citep{Gregory2012} may naturally account for the dependence of obliquities upon stellar (main sequence) $T_{\textrm{eff}}$. Crucially, if the star is inclined relative to its planetary system whilst young, fast-rotating and expanded \citep{Shu1987,Bouvier2013}, its quadrupole moment can be large enough to perturb a coplanar system of planets into a mutually-inclined configuration after disk dissipation. 

In what follows, we analyze this process quantitatively. First, we calculate the mutual inclination induced between two planets as a function of stellar oblateness ($J_2$), demonstrating a proof-of-concept that stellar obliquity suffices as a mechanism for over-producing single transiting systems. Following this, we use N-body simulations to subject the famed, 6-transiting system \textit{Kepler}-11 to the quadrupole moment of a tilted, oblate star. We show that not only are the planetary orbits mutually inclined, but for nominal parameters the system itself can undergo a dynamical instability, losing 3-5 of its planets, with larger mass planets preferentially retained. In this way, we naturally account for the slightly larger observed size of singles \citep{Johansen2012}.

\section{Analytical Theory}

In order to motivate the following discussion, consider two planets, orbiting in a shared plane around an inclined, oblate (high $J_2$) star. The effect of the stellar potential is to force a precession of each planetary orbit about the stellar spin pole, with the precession rate higher for the inner planet. If planet-planet coupling is negligible, the subsequent evolution would excite a mutual inclination between the planets of twice the stellar inclination (assuming fixed stellar orientation and negligible eccentricities). Alternatively, if planet-planet coupling is very strong, they will retain approximate coplanarity. Below, we analytically compute the system's evolution between these two extreme regimes (i.e., for general $J_2$). 

\subsection{Assumptions}

We restrict our analytic calculation to small mutual inclinations between the planets and utilize Laplace-Lagrange secular theory \citep{Murray1999}. This framework assumes the planets to be far from mean motion resonances, allowing one to average over the orbital motion. Consequently, each planetary orbit becomes dynamically equivalent to a massive wire, a concept that is due to Gauss \citep{Murray1999,Morby2002}. Furthermore we set all eccentricities to zero\footnote{This approximation is simply for ease of analytics and will be lifted in the numerical analysis below.}. 

The star's orientation will be held fixed. The validity of this assumption can be demonstrated by considering the ratio of stellar spin to planetary orbital angular momenta:
\begin{align}
\frac{J_\star}{\Lambda_{\textrm{p}}}&\equiv\frac{I_\star M_\star R_\star^2 \Omega_\star}{m_p\sqrt{G M_\star a_p}}
\end{align}
where $I_\star\approx0.21$ is the dimensionless moment of inertia appropriate for a fully convective, polytropic star \citep{Chandrasekar1939}, and the stellar rotation rate is $\Omega_\star=2\pi/P_\star$. Consider a young, Sun-like star, possessing a rotation period of $P_\star=10\,$days (on the slower end of observations; \citealt{Bouvier2013}) and a radius of roughly $2R_\odot$ \citep{Shu1987}. A 10 Earth-mass object would need to orbit at over $\sim$100\,AU in order to possess the angular momentum of the star.  Thus, provided we deal with compact, relatively low-mass systems, the stellar orientation can be safely fixed to zero.

A further assumption is that the dynamical influence of stellar oblateness may be approximated using only the leading order quadrupole terms, neglecting those of order $\mathcal{O}(J_2^2)$. Therefore, the disturbing part of the stellar potential (with $e=0$) may be written as \citep{Danby1992}
\begin{align}\label{disturbed}
\mathcal{R}&=\frac{G m_{\textrm{p}} M_\star}{2a_p}\bigg(\frac{R_\star}{a_p}\bigg)^2J_2\bigg(\frac{3}{2}\sin^2 i_p-1\bigg)\nonumber\\
&\approx\frac{G m_{\textrm{p}} M_\star}{2a_p}J_2\bigg(\frac{R_\star}{a_p}\bigg)^2\bigg(6s_p^2-1\bigg),
\end{align}
where the second step has made the assumption of small planetary inclination $i_p$ and defined a new variable $s_p\equiv\sin(i_p/2)$ (this definition is introduced to maintain coherence with traditional notation in celestial mechanics, e.g. \citet{Murray1999}).

Finally, it is essential to define an initial configuration for the planetary system. Both numerical and analytic modeling of planet-disk interactions suggest that embedded protoplanets have their inclinations and eccentricities damped to small values within the disk-hosting stage \citep{Tanaka2004,Cresswell2007,Kley2012}. Furthermore, any warping of the disk in response to a stellar companion is expected to be small \citep{Fragner2010}. Therefore, throughout this work we assume that the planets emerge from the disk with circular, coplanar orbits that are inclined by some angle $\beta_\star$ relative to the star. Note that we will always fix the stellar spin direction to be aligned with the $z$-axis, so $\beta_\star$, the stellar obliquity, constitutes the initial inclination of the planetary orbits in our chosen frame. 

\subsection{2-planet system}

Incorporating the above assumptions, we may now write down the Hamiltonian ($\mathcal{H}$) that governs the dynamical evolution of the planetary orbits. To second order in inclinations (and dropping constant terms) we have \citep{Murray1999} 
\begin{align}
\mathcal{H}=\underbrace{\frac{G m_1 m_2}{a_2}\bigg[\big(s_1^2+s_2^2\big)f_3+s_1s_2f_{14}\cos(\Omega_1-\Omega_2)\bigg]}_{\textrm{Planet-planet interaction}}\nonumber\\
\underbrace{-\frac{3G m_1 M_\star}{a_1}J_2\bigg(\frac{R_\star}{a_1}\bigg)^2s_1^2-\frac{3G m_2 M_\star}{a_2}J_2\bigg(\frac{R_\star}{a_2}\bigg)^2s_2^2}_{\textrm{Planet-stellar quadrupole interaction}},
\end{align}
where the prefactors are
\begin{align}
f_3=-\frac{1}{2}f_{14}=-\frac{1}{2}\bigg(\frac{a_1}{a_2}\bigg)b_{3/2}^{(1)}\bigg(\frac{a_1}{a_2}\bigg),
\end{align}
and $b_{3/2}^{(1)}$ is the Laplace coefficient
\begin{align}
b_{3/2}^{(1)}(\alpha)\equiv\frac{1}{\pi}\int^{2\pi}_0\bigg[\frac{\cos \psi}{(1+\alpha^2-2\alpha\cos \psi)^{3/2}}\bigg]d\psi.
\end{align}

As we are using Hamiltonian mechanics, the dynamics must be described in terms of canonical variables. Traditional Keplerian orbital elements do not constitute a canonical set, so we transform to Poincar\'e (or, modified Delauney; \citealt{Murray1999}) variables, defined as
\begin{align}
Z_{\textrm{p}}&\equiv m_{\textrm{p}} \sqrt{G M_\star a_{\textrm{p}}}\big(1-\cos(i_{\textrm{p}})\big)\,\,\,\,\,\,z_{\textrm{p}}\equiv-\Omega_{\textrm{p}}.
\end{align}
Physically, $Z_{\textrm{p}}$ is the angular momentum of a circular orbit after subtracting its component in the $z$-direction. Notice that in the small angle limit,
\begin{align}
Z_{\textrm{p}}\approx \frac{1}{2}m_{\textrm{p}} \sqrt{G M_\star a_{\textrm{p}}}i_{\textrm{p}}^2\equiv\frac{1}{2}\Lambda_{\textrm{p}}i_{\textrm{p}}^2.
\end{align}

 After substituting, we arrive at the governing Hamiltonian 
\begin{align}
\mathcal{H}=&-\mathcal{E}\bigg[\frac{Z_1}{\Lambda_1}+\frac{Z_2}{\Lambda_2}-2\sqrt{\frac{Z_1Z_2}{\Lambda_1\Lambda_2}}\cos(z_1-z_2)\bigg]\nonumber\\
&-\frac{3}{2}n_1J_2\bigg(\frac{R_\star}{a_1}\bigg)^2Z_1-\frac{3}{2}n_2J_2\bigg(\frac{R_\star}{a_2}\bigg)^2Z_2
\end{align}
where for compactness we define
\begin{align}
\mathcal{E}\equiv\frac{G m_1 m_2}{4a_2}\bigg(\frac{a_1}{a_2}\bigg)b_{3/2}^{(1)}\bigg(\frac{a_1}{a_2}\bigg).
\end{align}

In order to complete the calculation, we define a complex variable for that represents the inclination of each planet
\begin{align}
\eta_{\textrm{p}}&\equiv \sqrt{\frac{Z_{\textrm{p}}}{\Lambda_{\textrm{p}}}}\big(\cos(z_{\textrm{p}})+\imath\sin(z_{\textrm{p}})\big)\nonumber\\
&\approx\frac{1}{\sqrt{2}}i_{\textrm{p}}\big(\cos(\Omega_{\textrm{p}})-\imath\sin(\Omega_{\textrm{p}})\big),
\end{align}
where $\imath=\sqrt{-1}$. The purpose is to cast Hamilton's equations into an eigenvector/eigenvalue problem. Specifically, in terms of these new variables, we must solve
 \begin{align}
 \dot{\eta}_{\textrm{p}}=\imath\frac{\partial \mathcal{H}}{\partial \eta_{\textrm{p}}^*}\frac{1}{\Lambda_{\textrm{p}}},
 \end{align}
in which ``$\,^*\,$" denotes complex conjugation, yielding the matrix equation
\[\frac{d}{dt}
\begin{pmatrix}
    \eta_1  \\
    \eta_2  \\
\end{pmatrix}
=-\imath 
\begin{pmatrix}
    B_1+\nu_1& -B_1  \\
   -B_2& B_2+\nu_2  \\
\end{pmatrix}
\begin{pmatrix}
    \eta_1  \\
    \eta_2  \\
\end{pmatrix},
\]
where we have defined four frequencies as
\begin{align}
B_1\equiv\frac{1}{4}n_1 \bigg(\frac{a_1}{a_2}\bigg)^2b_{3/2}^{(1)}\bigg(\frac{a_1}{a_2}\bigg)\frac{m_2}{M_\star}\nonumber\\
B_2\equiv\frac{1}{4}n_2 \bigg(\frac{a_1}{a_2}\bigg)b_{3/2}^{(1)}\bigg(\frac{a_1}{a_2}\bigg)\frac{m_1}{M_\star}\nonumber\\
\nu_{\textrm{p}}=\frac{3}{2}n_{\textrm{p}}J_2\bigg(\frac{R_\star}{a_{\textrm{p}}}\bigg)^2.
\end{align}

The equation above may be solved using standard methods, whereby the solution is written as a sum of eigenmodes
\begin{align}
\eta_{\textrm{p}}=\sum_{j=1}^2\eta_{\textrm{p},j}\exp(\imath \lambda_j t).
\end{align}
 Indeed, the problem may be easily extended to $N$ planets, though writing down all eigenvectors $\eta_{\textrm{p},j}$ and eigenmodes $\lambda_j$ rapidly becomes cumbersome.  

\subsubsection{Initial conditions and solution}
A choice must be made for the initial conditions of the problem. As already mentioned above, here we choose the condition that both orbits are initially coplanar, having recently emerged from their natal disk, with the star inclined by some angle $\beta_\star$ relative to them. Accordingly, all four boundary conditions may be satisfied by requiring that
\begin{align}
\eta_{\textrm{p}}\big|_{t=0}=\frac{\beta_\star}{\sqrt{2}}.
\end{align}

What we seek is the mutual planet-planet inclination (denoted $\beta_{\textrm{rel}}$). In the small angle approximation, we can compute this quantity using the relation
\begin{align}\label{incline}
(1-\cos(\beta_{\textrm{rel}}))&\approx\eta_1\eta_1^*+\eta_2\eta_2^*-(\eta_1\eta_2^*+\eta_2\eta_1^*)\nonumber\\
&\approx\frac{1}{2}\beta_{\textrm{rel}}^2.
\end{align}

\begin{figure}[h!]
\centering
\includegraphics[trim=4cm 6cm 2cm 3cm, clip=true,width=1.2\columnwidth]{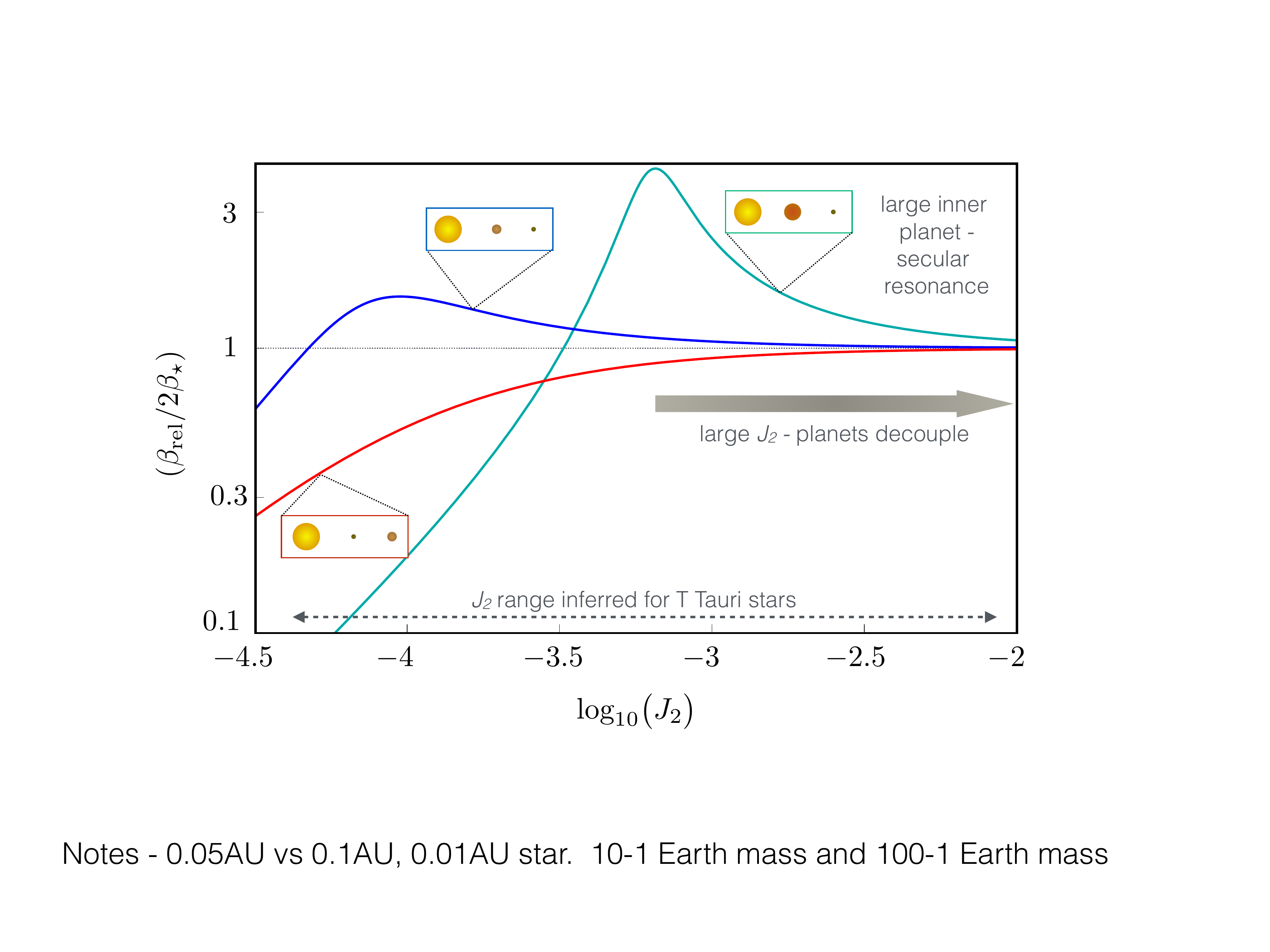}
\caption{The amplitude of oscillations in mutual planet-planet inclinations excited between two initially coplanar, circular planetary orbits $\beta_{\textrm{rel}}$, scaled by twice the stellar obliquity $\beta_\star$. The planets are situated at 0.05\,AU and 0.1\,AU for 3 different mass configurations: The red line has a 10 Earth mass planet outside a 1 Earth mass planet, where blue has the planets switched. The cyan line augments the inner planet to 100 Earth masses. Notice that any time the inner planet has more angular momentum, there exists a peak in the misalignments, representing resonance. In the limit of large $J_2$, the planets entirely decouple and reach mutual inclinations equal to twice the stellar obliquity.}
\label{analytic}
\end{figure}

After solving for eigenvalues, eigenvectors and matching the initial conditions, we arrive at the solution for the mutual inclination of the two planets as a function of time, which takes the rather compact form
\begin{align}\label{analyticEq}
\boxed{\beta_{\textrm{rel}}(t)=2\beta_\star \mathcal{G}(J_2)\sin(\omega_0t/2)}
\end{align}
where we define the (semi-)amplitude of the oscillations between planets 
\begin{align}
\mathcal{G}=\mathcal{L}\bigg[1+\mathcal{L}^2+2\bigg(\frac{\Lambda_2-\Lambda_1}{\Lambda_2+\Lambda_1}\bigg)\mathcal{L}\bigg]^{-1/2}
\end{align}
in terms of the ratio of frequencies
\begin{align}
\mathcal{L}&\equiv \frac{\nu_1-\nu_2}{B_1+B_2}\nonumber\\
&=6J_2\bigg(\frac{R_\star}{a_1}\bigg)^2\bigg(\frac{a_2}{a_1}\bigg)^2 \frac{1}{b_{3/2}^{(1)}(\alpha)}\frac{M_\star}{m_2}\frac{1-\alpha^{7/2}}{1+\Lambda_1/\Lambda_2}.
\end{align}
A convenient consequence of the aligned initial conditions is that the oscillations are purely sinusoidal, evolving with half the frequency 
\begin{align}
\omega_0&\equiv\big[(B_1+B_2)^2+(\nu_1-\nu_2)\nonumber\\
&\times\big(\nu_1-\nu_2+2(B_1-B_2)\big)\big]^{1/2}.
\end{align}

One aspect to notice is that the amplitude is maximized when the equality
\begin{align}
\nu_1+\mathcal{K}^2 B_1=\nu_2+\mathcal{K}^2 B_2
\end{align}
is satisfied, where $\mathcal{K}\equiv (\Lambda_1+\Lambda_2)/(\Lambda_1-\Lambda_2)$. The maximum amplitude, scaled by $\beta_\star$, can then be written as
\begin{align}\label{Gmax}
2\mathcal{G}_{\textrm{max}}=\frac{\Lambda_1+\Lambda_2}{\sqrt{\Lambda_1\Lambda_2}}.
\end{align}

The significance of this result is best seen upon considering the outer planet to be a test particle, such that $B_1=0$ and $\Lambda_2\rightarrow0$. In such a scenario, the maximum of the amplitude $\mathcal{G}_{\textrm{max}}\rightarrow \infty$. Such an unphysical result occurs as a consequence of a secular resonance \citep{Murray1999,Morby2002,Spalding2014a,Batygin2015}, whereby the inner and outer bodies precess at similar rates. In reality, our earlier approximation that mutual inclinations are small breaks down in this regime and the inclusion of higher order terms is required. 

In principle, one may also set $\Lambda_1\rightarrow0$ and conclude that the above resonance persists when the inner planet is a test particle. However, the resonant criterion in terms of stellar oblateness reads:
\begin{align}\label{res}
J_2\big|_{\textrm{res}}&\approx \frac{1}{6}\bigg(\frac{a_1}{R_\star}\bigg)^2\bigg(\frac{a_1}{a_2}\bigg)^2\bigg(\frac{m_2}{M_\star}\bigg)\nonumber\\
&\times \Bigg(\frac{b_{3/2}^{(1)}(\alpha)}{\Lambda_2}\Bigg)\frac{(\Lambda_1+\Lambda_2)^2}{(\Lambda_1-\Lambda_2)(1-\alpha^{7/2})},
\end{align}
which is negative when $\Lambda_1<\Lambda_2$. Accordingly, the condition for secular resonance can only be satisfied when $\Lambda_1>\Lambda_2$, i.e., when the inner planet possess more orbital angular momentum than the outer planet. 

As an illustration, we plot the semi-amplitude $\mathcal{G}$ in Figure~(\ref{analytic}) appropriate for a configuration where the two planetary orbits are situated at 0.05\,AU and 0.1\,AU, both orbiting a solar-mass star with radius $0.01$\,AU (about twice the Sun's radius). Three cases are shown: The red line depicts a 1 Earth mass planet interior to a 10 Earth mass planet while the blue line has the planets interchanged. The former configuration possesses a positive $J_2\big|_{\textrm{res}}$, appearing as a maximum in the amplitude. The third situation (the cyan line) represents a 100 Earth mass planet interior to a 1 Earth mass body, illustrating that higher mass inner planets may more easily misalign their outer companion (equation~\ref{Gmax}), but $J_2\big|_{\textrm{res}}$ is correspondingly higher. 

Figure~(\ref{analytic}) demonstrates that misalignments of the order twice the stellar obliquity can be readily excited for reasonable values of $J_2$. By geometric arguments, the potential for such misalignments to take one of the planets out of transit depends upon the ratio $R_{\star}/a$. However, for the cases considered above, only $\sim4^\circ$ of stellar obliquity are required to remove the two planets from a co-transiting configuration (less than the $\sim7^\circ$ present in the solar system; \citealt{Lissauer2011b}). Conversely, planets may remain co-transiting if the innermost planet is sufficiently distant, the planets are very massive and/or tightly packed, or the stellar quadrupole moment is particularly low. 

\section{Numerical Analysis}

Several crucial aspects of real systems were neglected in order to obtain the analytic solution~(\ref{analyticEq}) above. Principally, we included only two planets whose orbits were assumed to be circular and only slightly mutually inclined. Additionally, in averaging over short-term motion, our adopted secular approach is unable to describe the full dynamics. A more subtle aspect was that we considered a constant $J_2$, when in reality, stars are expected to spin down and shrink over time until $J_2$ is essentially negligible \citep{Irwin2008,McQuillan2013,Bouvier2013}. 

In order to test our hypothesis within a more general framework, we now turn to N-body simulations. To carry out the calculations, we employed the well-tested Mercury6 symplectic integration software package (\citealt{Chambers1999}) \footnote{http://www.arm.ac.uk/~jec/home.html}. In addition to standard planet-planet interactions, we modified the code to include the gravitational potential arising from a tilted star of given $J_2$, along with a term to produce general relativistic precession (following \citet{Nobili1986}).  

For the sake of definiteness, the parameters of our modelled system were based off of \textit{Kepler}-11, a star around which 6 transiting planets have been discovered \citep{Lissauer2011a}. Detailed follow-up studies, using Transit Timing Variations, have constrained the masses of the innermost 5 planets and placed upper limits upon the mass of \textit{Kepler}-11g, the outermost planet\footnote{The mass of \textit{Kepler}-11g, is only loosely constrained, however for the purposes of this work, it is not particularly imperative to choose the ``real" mass.}, making this system ideal for dynamical investigation. Though choosing one system is not exhaustive, our goal is to demonstrate the influence of a tilted star upon a general coplanar system of planets. We follow \citet{Lissauer2013} and use their best-fit mass of $8$\,Earth Masses for \textit{Kepler}-11g, with the stellar mass given by $0.961M_\odot$ (see Table~\ref{Kepler11}).

 \subsection{N-body simulation}
 
 For our numerical runs, we choose 10 values of stellar obliquity and 11 of initial stellar $J_2=J_{2,0}$ (i.e., the oblateness immediately as the disk dissipates). Once again, we fix the stellar orientation aligned with the $z$-axis, but choose the initial planet-star misalignments:
 \begin{align}
 \beta_\star \in\{5,\,10,\, 20,\, 30,\, 40,\, 50,\, 60,\, 70,\, 80,\, 85\}.
 \end{align}
  The value of $J_2$ for a star deformed by its own rotation may be related to its spin rate $\omega_\star$ and Love number (twice the apsidal motion constant) $k_2$ as follows \citep{Sterne1939}:
\begin{align}
J_2=\frac{1}{3}\bigg(\frac{\Omega}{\Omega_b}\bigg)^2\,k_2,
\end{align}
where $\Omega_b$ is the break-up spin frequency at the relevant epoch. The Love number can be estimated by modeling the star as a polytope with index $\chi=3/2$ (i.e., fully convective; \citealt{Chandrasekar1939}), which yields $k_2\approx0.28$. Observations constrain the spin-periods of T-Tauri stars to lie within the range $\sim1-10$\,days \citep{Bouvier2013}, while the break-up period is given by
\begin{align}
T_\textrm{b}=\frac{2\pi}{\Omega_b}\approx\frac{1}{3}\,\bigg(\frac{M_\star}{M_\odot}\bigg)^{-1/2}\bigg(\frac{R_\star}{2R_\odot}\bigg)^{3/2}\textrm{days}.
\end{align}

In our simulations below, we use the current mass of \textit{Kepler}-11 for the star, but suppose its radius to be somewhat inflated relative to its current radius ($R_\star=2R_\odot$), reflecting the T Tauri stage \citep{Shu1987}. With these parameters, we arrive at a reasonable range of $J_{2,0}$ of
\begin{align}
10^{-4}\lesssim J_{2,0} \lesssim 10^{-2},
\end{align}
within which we choose 11 values uniformly separated in log-space:
\begin{align}
 J_{2,0}\in\{10^{-4},\, 10^{-3.8}\,...\,10^{-2}\}.
 \end{align}

   In all runs, rather than allowing both $R_\star$ and $J_2$ to vary, we simply left $R_\star$ as a constant, letting $J_2$ decay exponentially over a timescale of $\tau=1$\,Myr
\begin{align}
J_2(t)=J_{2,0}\,\textrm{e}^{(-t/\tau)}.
\end{align}
The choice for $\tau$ is essentially arbitrary, provided $J_2$ decays over many precessional timescales, owing to the adiabatic nature of the dynamics \citep{Lichtenberg1992,Morby2002}. Our choice of 1\,Myr roughly coincides with a Kelvin-Helmholtz timescale \citep{Batygin2013} but is chosen also to save computational time. 

For each case, our integrations span 22 million years, beginning with the initial condition of a coplanar system possessing the current semi-major axes of the \textit{Kepler}-11 system, but with eccentricities set to zero \citep{Lissauer2013}. In order to analyze the results, we sample the system 6 times between 19 and 22 Myr, and at each step calculate the maximum number of transiting planets that could be observed from a single direction. The results at all six times were then averaged.
 
The determination of the maximum number of transits was accomplished as follows. We begin by checking whether all possible pairs within the 6 planets mutually transit, where the criterion for concluding a pair of planets to be non-transiting is: 
\begin{align}
\big|\sin(\beta_{\textrm{rel}})\big|>\sin(\beta_{\textrm{crit}})\approx \frac{R_{\star}}{a_1}+\frac{R_{\star}}{a_2},
\end{align}
where in the above formula we used the current radius of \textit{Kepler}-11 ($R_\star=1.065R_\odot$, as opposed to the inflated value relevant to the T Tauri stage). If any single pair of planets was non-transiting, we proceeded to choose each possible combination of 5 out of the 6 and performed a similar pairwise test to identify potentially observable 5-transiting systems. If no set of 5 passed the test we chose all sets of 4, etc, until potentially finding that only one planet could be seen transiting. We note that the criterion above neglects the possibility of fortunate orbital configurations allowing two mutually inclined orbits to intersect along the line of sight. This complication, however, does not affect our qualitative picture.

\section{Results}

\begin{figure}[h!]
\centering
\includegraphics[trim=0cm 0cm 0cm 0cm, clip=true,width=1\columnwidth]{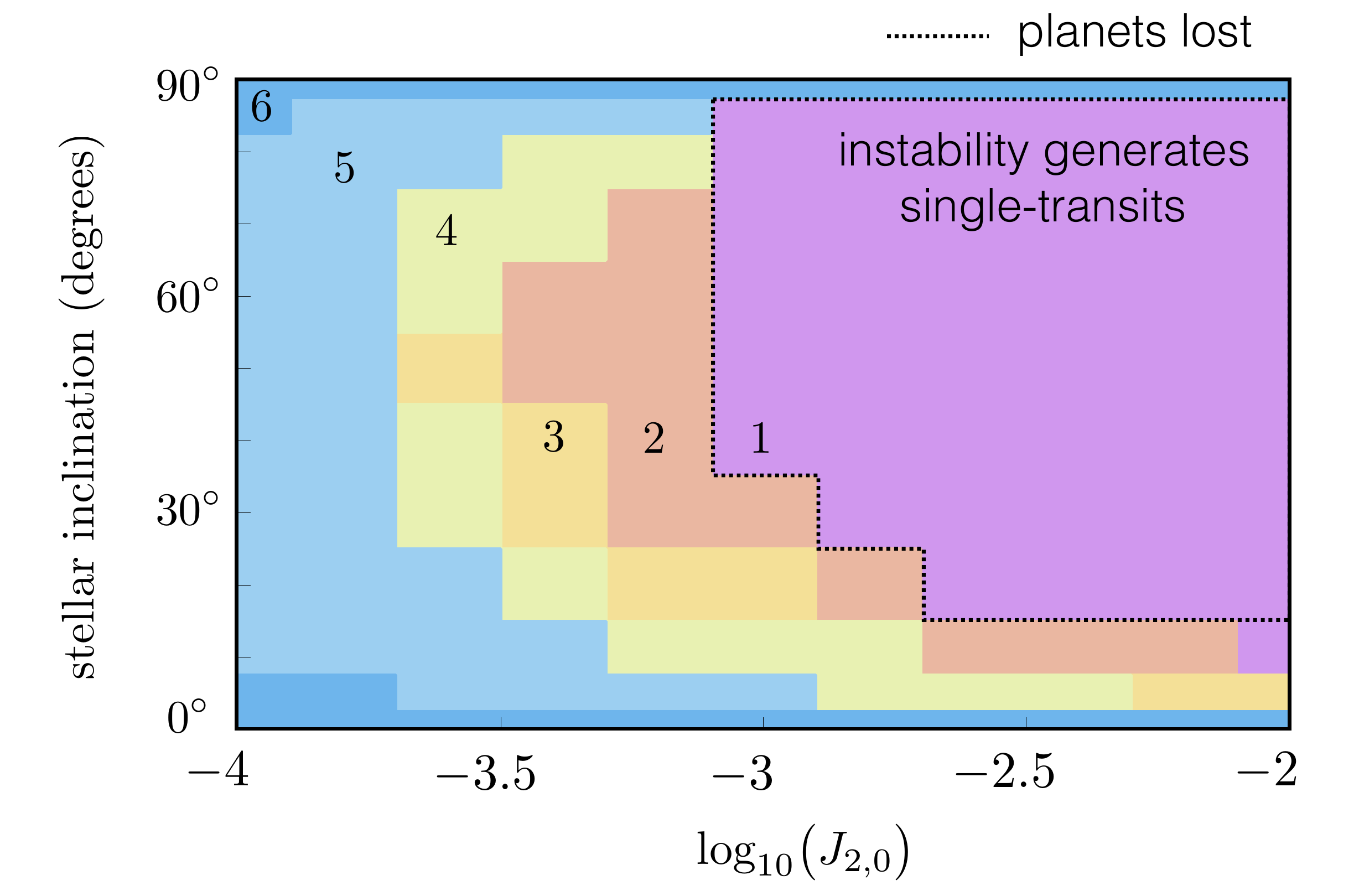}
\caption{The maximum number of transits detectable after 22 million years of integrating \textit{Kepler}-11 with a tilted, oblate star. The $x$-axis denotes the value of $J_2$ immediately after disk dispersal ($J_{2,0}$) and the $y$-axis represents the stellar inclination. The runs where planets were lost through instability are outlined by a dotted line, which corresponds closely to the region where only single transits can be observed (the purple region).}
\label{Numerical}
\end{figure}

The numerical results are presented in Figure~(\ref{Numerical}), where the $x$-axis depicts the initial stellar $J_2=J_{2,0}$. The $y$-axis refers to the misalignment $\beta_\star$ between the stellar spin axis and the initial plane of the 6 planets. Each run has been given its own rectangular box, within which the color represents the maximum number of co-transiting planets that may be observed around the star (as discussed above). The numerics verify our analytic result, in that the observable multiplicity may be significantly reduced solely as a consequence of stellar obliquity. 

As expected, higher values of $J_{2,0}$ result in fewer transiting planets, provided the star is tilted relative to the planetary orbits. As with our analytic results, even small stellar obliquities are sufficient to reduce the transit count, with 5$^\circ$ of obliquity reducing the transit number to as little as 3 (Figure~(\ref{Numerical})). However, planet-planet mutual inclination was not the only source of the reduction in transit number. A crucial finding was that for large enough $J_{2,0}$ and $\beta_\star$, the stellar quadrupole potential caused the system to go unstable, casting 3-5 planets out of the system or into the central body, with planet-planet collisions existing as an additional possibility not captured in our simulations \citep{Boley2016}. 

 The region of instability (i.e., where at least one planet was lost) is outlined by a dotted line in Figure~(\ref{Numerical}). Interestingly, the areas of instability map closely onto the regions where only a single transit remains. In other words, almost every single-transiting system coming out of the integration had lost planets through dynamical instability. Furthermore, each time instability occurred, the 2 lowest mass members were lost: \textit{Kepler}-11b and f, with the next lowest mass body, \textit{Kepler}-11c often joining them. Such a preference for retaining more massive planets is indeed reflected in the data as a slightly larger typical radius for single-transiting systems \citep{Johansen2012}. More testing is required to determine whether this a generic feature of our model, however.

We showed that for any given two-planet system, there exists a resonant $J_2$ if the inner planet has more angular momentum than the outer. However, the picture becomes much more complicated in a multi-planet system, where each planet introduces two additional secular modes (one for eccentricity one for inclination; \citealt{Murray1999}), increasing the density of resonances in Fourier space. As the stellar $J_2$ decays, its influence sweeps across each resonance, providing ample opportunity to excite mutual inclinations. If the two planets, \textit{Kepler}-11d and f, were alone, they could become resonant at $J_2\big|_{\textrm{res}}\sim10^{-2.4}$, which coincides approximately with the onset of instability in the low-inclination runs (Figure~\ref{Numerical}) but not exactly, for the reasons mentioned above.

 \section{Discussion}

In recent years, the \textit{Kepler} dataset has grown sufficiently comprehensive to facilitate statistically robust investigations at a population level. Out of this data has emerged a  so-called ``\textit{Kepler} Dichotomy"; the notion that single-transiting systems are too common to be explained as resulting from a simple distribution of mutual inclinations within systems of higher multiplicity \citep{Johansen2012,Ballard2016}.

 In separate studies, significant misalignments have been detected between the stellar spin axes and the planetary orbits they host, particularly around stars with main sequence $T_{\textrm{eff}}\gtrsim6200$\,K \citep{Winn2010,Albrecht2012,Mazeh2015,Li2016}. This trend initially became apparent within the hot Jupiter dataset and was consequently often interpreted as evidence for a post-disk, high-eccentricity migration pathway for hot Jupiter formation\footnote{The dependence on host star temperature (and therefore mass) was attributed to tidal dissipation within the convective regions of lower-mass stars \citep{Winn2010,Lai2012}.}. However, a similar trend has now emerged within \textit{Kepler} systems, including the multi-transiting sub-population \citep{Huber2013,Mazeh2015}, with little evidence supporting a tidal origin \citep{Li2016}. These observations cumulatively suggest that many of the misalignments originated from directly tilting the protoplanetary disk, thereby inclining all planets in the system at once \citep{Batygin2012,Spalding2014a,Lai2014,Spalding2015,Fielding2015}. 
 
 A consequence of primordially-generated spin-orbit misalignments is that stellar obliquity would be present at the end of the natal disk's life, leaving the planetary orbital architecture subject to the quadrupole moment of their young, rapidly-spinning and expanded host star. This paper has demonstrated that such a configuration naturally misaligns close-in systems and, furthermore, provides a mechanism for dynamical instability that by-passes the problem encountered in earlier work that unreasonably large planets were required to induce instability \citep{Johansen2012}.
   
 The observable multiplicity of transiting systems can be reduced either by inclining planetary orbits relative to each other, or by intrinsically reducing the number of planets. Here, we have shown that both can be at play, with modest $J_2$ and stellar obliquity causing misalignments, whereas sufficiently large values thereof lead to dynamical instability, shedding planets. The origin of the instability is likely secular in nature, and significant planet-planet inclinations have been shown to reduce the inherent stability of planetary systems in numerous previous works \citep{Laughlin2002,Veras2004,Nelson2014}. In support of this interpretation, our simulations resulted in planetary instability at much smaller $J_2$ when obliquity was high $\gtrsim40^\circ$. Accordingly, we would expect multiplicity (both transiting and intrinsic) to be lower around hot stars, which tend to possess higher obliquities (\citealt{Winn2010}; see below).

 \subsection{Predictions}
 
 Imposing stellar obliquity as a source of the \textit{Kepler} Dichotomy leads to several predictions. Naturally, stars leaving the disk-hosting stage with larger $J_2$ and obliquity are more likely to end up observed as exhibiting single-transits, either as a result of dynamical instability or the excitation of mutual planet-planet inclinations. As mentioned above, there is an observed trend whereby stars with $T_{\textrm{eff}}\gtrsim6200\,$K exhibit higher obliquity \citep{Winn2010,Albrecht2012,Mazeh2015} and so, on the face of it, one would expect a higher relative incidence of singles around higher mass stars. The picture is, however, complicated by the universal feature of stellar evolution models that more massive stars contract along their Hayashi tracks faster \citep{Siess2000}. Accordingly, the influence of $J_2$ in more massive stars may have decayed to a greater extent than in lower-mass stars by the time their natal disk dissipates, partly offsetting the impact of their larger typical obliquities. 
 
 The above complications notwithstanding, both our analytical and numerical analyses suggest a greater sensitivity of the degree of misalignment to stellar obliquity than to stellar $J_2$. Consequently, we make the prediction that \textit{hot stars possess more abundant single-transiting systems relative to cool stars}.
  
 Previous work has already suggested the existence of our predicted trend. Specifically, both hotter stars and, independently, single-transiting systems appear to exhibit higher obliquities \citep{Morton2014,Mazeh2015}. The overlapping of these two findings implies at least a weak trend toward more singles around hotter stars. In order to further test this prediction, we carried out a simple statistical analysis of confirmed \textit{Kepler} planets, as we now describe. 
 
 \subsection{\textit{Kepler} data}
To obtain data on confirmed, \textit{Kepler} systems, we downloaded the data from the ``Confirmed Planets" list (as of July 2016) on the NASA Exoplanet Archive website\footnote{http://exoplanetarchive.ipac.caltech.edu}. The systems were filtered to include only those in the \textit{Kepler} field, though the conclusions that follow do not change qualitatively if non-\textit{Kepler} detections are included. 
 
\begin{figure}[h!]
\centering
\includegraphics[trim=0cm 0cm 0cm 0cm, clip=true,width=1\columnwidth]{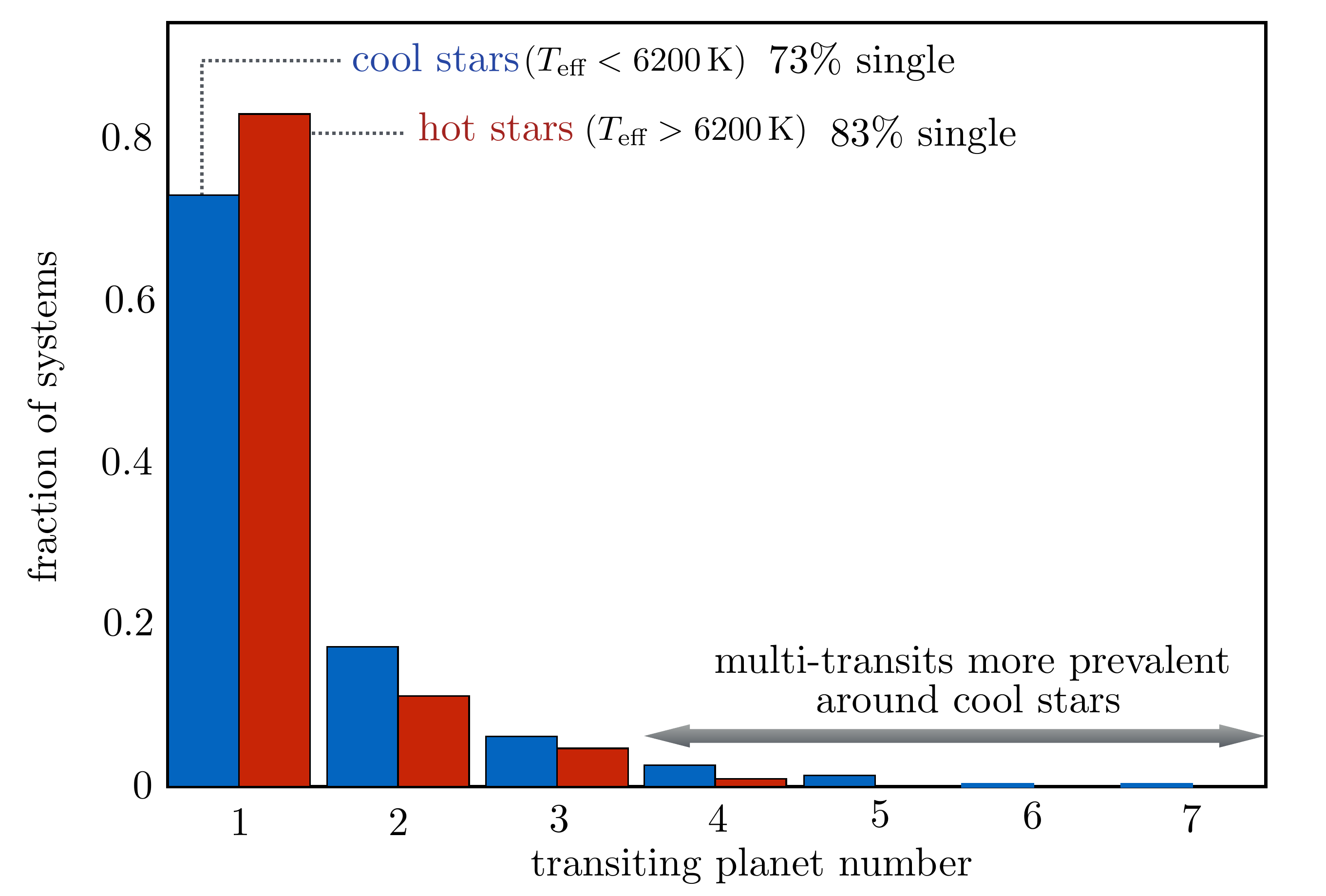}
\caption{Fraction of systems exhibiting each number of transiting planets from 1 to 7 within the hot ($T_{\textrm{eff}}>6200$\,K, red bars) and cool ($T_{\textrm{eff}}<6200$\,K, blue bars) sub-samples of planet-hosting Kepler stars. There were 132 hot stars and 1504 cool stars in the data used, of which 83\% and 73\% respectively exhibited single transits. Accordingly, transiting systems around hot stars show a stronger tendency toward being single, in agreement with the predictions of our presented model (see text).}
\label{Chart}
\end{figure}

In Figure~(\ref{Chart}), we split the data into ``hot" stars with $T_{\textrm{eff}}>6200\,$K (132 in total) and ``cool" stars, with $T_{\textrm{eff}}<6200\,$K (1504 in total). For each sub-population, we illustrate the fraction of systems as a function of the number of planets observed in transit. The hot stars clearly demonstrate a larger fraction of singles and a smaller fraction of multiples for each value of multiplicity, in agreement with the predictions of our model. In order to quantify the significance of this agreement, we carry out a statistical test that quantitatively compares the two populations.
 
 Using Bayes' theorem, with a uniform prior, we generated a binomial distribution for hot stars and cool stars separately that illustrates the probability of the data, given an assumption about what fraction of systems are single. Such an argument is similar to determining the fairness of a coin flip, where heads equates to a single system and tails a multi system. Specifically, we plot
 \begin{align}
 \mathcal{P}(\{\textrm{data}\}|S)=A S^{N_\textrm{s}}(1-S)^{N_\textrm{t}-N_\textrm{s}},
 \end{align} 
where $N_\textrm{s}$ is the number of single systems within a population of $N_\textrm{t}$ total stars and $A$ is a normalization coefficient \citep{Sivia1996}. The variable $S$ is the single bias weighting; the probabilistic tendency for a population of planet-hosting stars to display single transits as opposed to multiples. The quantity $\mathcal{P}(\{\textrm{data}\}|S)$ gives the probability of reproducing the data if the underlying tendency is $S$. In the hot population, $N_\textrm{t}=132$ and $N_\textrm{s}=110$, whereas the cool population had $N_\textrm{t}=1504$ and $N_\textrm{s}=1098$.

\begin{figure}[h!]
\centering
\includegraphics[trim=0cm 0cm 0cm 0cm, clip=true,width=1\columnwidth]{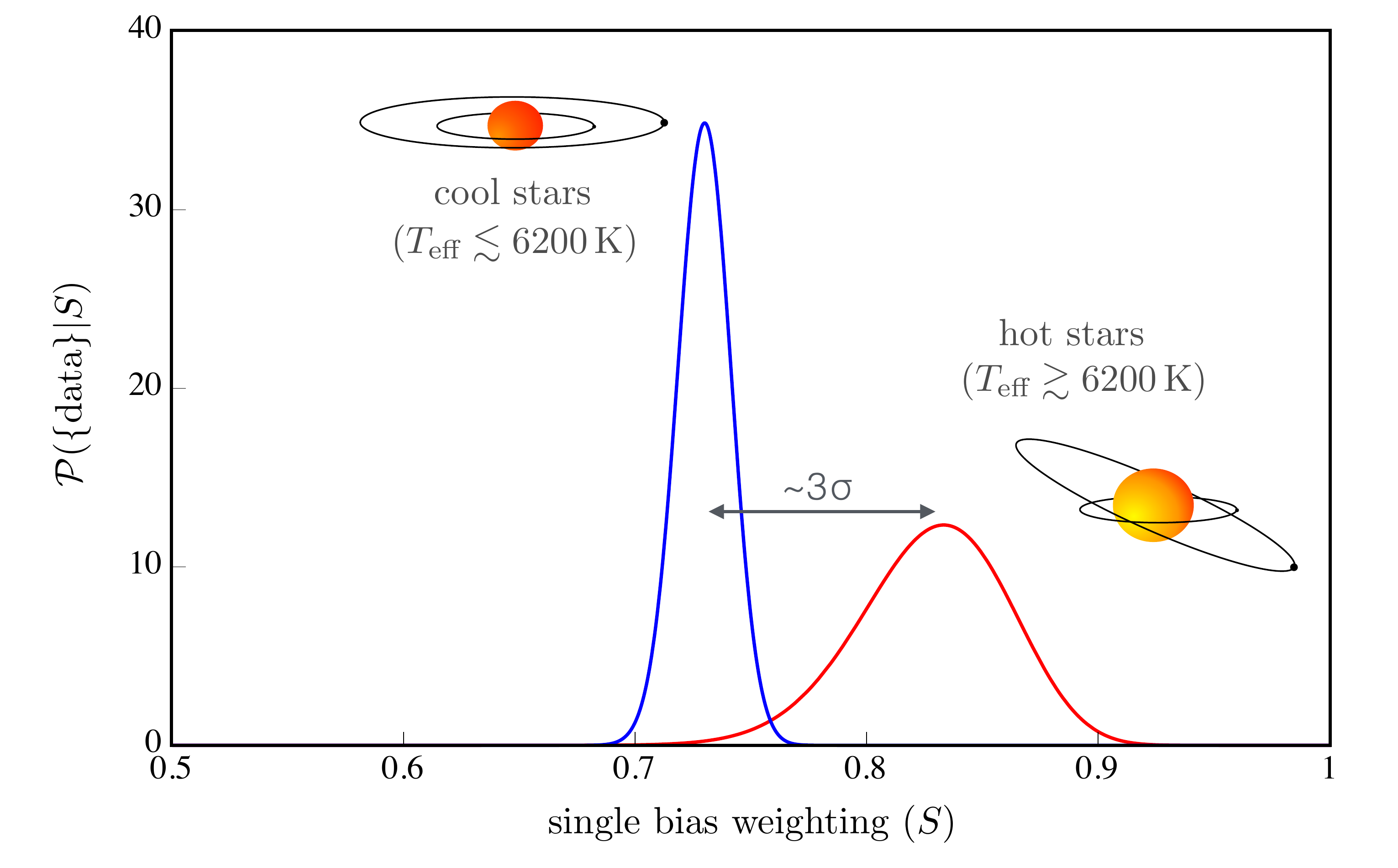}
\caption{Probability of the data, given an intrinsic fraction $S$ of singles out of systems with hot (red line) and cool (blue line) stars. The separation of the peaks is roughly 2.9\,$\sigma$ (as defined in the text). Therefore, to a very high confidence, hot stars possess relatively more singles, as our hypothesis predicts.}
\label{Stats}
\end{figure}

As can be seen from Figure~(\ref{Stats}), the two distributions are visually distinct, with hotter stars possessing a stronger bias towards singles, with a significance of $2.9\sigma$ \footnote{Where $\sigma^2$ here is defined as the sum of the squares of the standard deviations of each individual distribution.}. More data are needed to tease out this relationship further and to isolate the influence of a tilted star versus other confounding factors, such as the dependence upon stellar mass of the occurrence rate of giant planets. For now, we conclude that the data supports our general prediction, that hotter, more oblique stars possess a relatively greater abundance of single-transiting planets.

A separate prediction relates to the distance of the planets from the host star. Specifically, the quadrupole moment falls off as $R_\star^2/a^2$, whereas the coplanarity required for transit grows as $a/R_\star$, and so the overall magnitude of our proposed mechanism should become negligible within more distant systems. In consequence, we predict that the ``\textit{Kepler} Dichotomy" signal will weaken for systems at larger orbital distances. As future missions, such as \textit{TESS} collect more data, this unique aspect of our model will become amenable to observational tests.

A caveat to the above analysis is that the dichotomy appears to persist within the population of planets around M-dwarfs \citep{Ballard2016}. This is problematic as these stars are expected to exhibit lower inclinations, being cooler on the main sequence. We interpret this as stellar oblateness being effective across all stellar masses, but being relatively more important within the hotter, more inclined population. This is supported by our numerical simulations, where even small obliquities reduced the transit number if oblateness was large enough (Figure~(\ref{Numerical})). 

\subsection{The origin of spin-orbit misalignments}

Our work here essentially relies upon the assumption that stellar obliquity is excited early on, in the disk-hosting stage. This is not the only potential origin for spin-orbit misalignments, with alternative pathways including secular chaos \citep{Lithwick2012}, planet-planet scattering \citep{Ford2008} and Kozai interactions \citep{Naoz2011,Petrovich2015a}. These mechanisms are traditionally inseparable from the idea that hot Jupiters migrate through a post-disk, high-eccentricity pathway \citep{Wu2003,Petrovich2015b}. Whereas it is likely that some hot Jupiters formed in this way, it is unlikely to constitute the dominant pathway \citep{Dawson2014b} and, furthermore, cannot explain the spin-orbit misalignment distribution in \textit{Kepler} systems \citep{Mazeh2015,Li2016}. Rather, disk-driven migration constitutes a favourable mechanism that may retain multiple planet systems within the same plane, and can account for the observed spin-orbit misalignments if the disk itself becomes misaligned with respect to the host star.

To that end, multiple studies have shown that a stellar companion is dynamically capable of exciting star-disk misalignments across the entire observed range of spin-orbit misalignments \citep{Batygin2012,Spalding2014a,Lai2014}. Specifically, the tidal potential of a companion star, or even that of the star cluster itself, induces a precession of the disk orientation, leading to significant star-disk misalignments, usually by way of a secular resonance \citep{Spalding2014a}. Although observations of disk orientation in young binary systems are elusive, there exists at least one known example of a binary where each star has a disk with its plane misaligned to that of the binary \citep{Jensen2014}, just as in the aforementioned theoretical picture. Furthermore, stellar multiplicity appears to be a nearly universal outcome of star formation \citep{Duchene2013,Beuther2014}. 

Within the framework of primordial excitation of spin-orbit misalignments, the dependence upon stellar mass (or $T_{\textrm{eff}}$) has been linked to the observed multipolar field topology of higher mass T Tauri stars compared with the more dipolar configuration seen in lower-mass T Tauri stars \citep{Gregory2012}. The weaker dipoles of higher mass stars increase their magnetic realignment timescales above that of their lower-mass counterparts, naturally explaining the observed trend in spin-orbit misalignments with stellar $T_{\textrm{eff}}$, and therefore mass \citep{Spalding2015}. Our work here has demonstrated an additional consistency between observations and primordially excitied spin-orbit misalignments, namely that the \textit{Kepler} Dichotomy naturally arises from the dynamical response of multi-planet systems to the potential of an oblate, tilted star.

\section{Conclusions}
This paper investigates the origin of the ``\textit{Kepler} Dichotomy," within the context of primordially-generated spin-orbit misalignments. We have shown that the quadrupole moment of such misaligned, young, fast-rotating stars is typically capable of exciting significant mutual inclinations between the hosted planetary orbits. In turn, the number of planets available for observation through transit around such a star is reduced, either through dynamical instability or directly as a result of the mutual inclinations, leaving behind an abundance of single-transiting systems \citep{Johansen2012}. The outcome is an apparent reduction in multiplicity of tilted, hot stars, with their observed singles being slightly larger, as a consequence of many having undergone dynamical instabilities, in accordance with observations. 

Through the conclusions of this work, the origins of hot Jupiters, of compact \textit{Kepler} systems, the \textit{Kepler} Dichotomy and spin-orbit misalignments, are all placed within a common context.

\begin{table*}[t]
  \centering
\begin{tabular}{ |p{3cm}||p{1cm}|p{1cm}|p{1cm}|p{1cm}|p{1cm}|p{1cm}|  }
 \hline
 \multicolumn{7}{|c|}{Kepler-11} \\
 \hline
Property & b & c  & d & e & f & g\\
 \hline
 Mass (Earth masses) & 1.9 & 2.9 & 7.3 & 8.0 & 2.0 & 8.0\\
 Radius (Earth radii) & 1.80 & 2.87 & 3.12 & 4.19 & 2.49 & 3.33\\
 $a$ (AU) & 0.091 & 0.107 & 0.155 & 0.195 & 0.250 & 0.466\\
 Period (days) & 10.3 & 13.0 & 22.7 & 32.0 & 46.7 & 118.4 \\
 \hline
\end{tabular}  
 \caption{The parameters of the \textit{Kepler}-11 system. The mass of \textit{Kepler}-11g only has upper limits set upon it, but we follow \citet{Lissauer2013} and choose a best fit mass of 8\,Earth masses here.}
  \label{Kepler11}
\end{table*}
  
\begin{acknowledgments}
 This research is based in part upon work supported by NSF grant AST 1517936 and the NESSF Graduate Fellowship in Earth and Planetary Sciences (C.S). We would like to thank Erik Petigura, Henry Ngo and Peter Gao for helpful discussions, along with Fred Adams and Joe O'Rourke for useful suggestions. We are grateful to the anonymous reviewer for helpful comments that significantly improved the manuscript.
\end{acknowledgments}


\begin{thebibliography}

\bibitem[Albrecht et al.(2012)]{Albrecht2012}Albrecht, S., Winn, J. N., Johnson, J. A., Howard, A. W., Marcy, G. W., Butler, R. P., ... \& Hirano, T. (2012), ApJ, 757(1), 18.

\bibitem[Ballard \& Johnson(2016)]{Ballard2016}Ballard, S., \& Johnson, J., J. (2016), The Astrophysical Journal, 816, 66

\bibitem[Bate et al.(2010)]{Bate2010}Bate, M. R., Lodato, G., \& Pringle, J. E. 2010, \textit{MNRAS}, 401(3), 1505.

\bibitem[Batygin et al.(2011)]{Batygin2011}Batygin, K., Morbidelli, A., \& Tsiganis, K. (2011). A\&A, 533, 7.

\bibitem[Batygin(2012)]{Batygin2012}Batygin, K. 2012, Nature, 491(7424), 418.

\bibitem[Batygin \& Adams(2013)]{Batygin2013} 
Batygin, K., \& Adams, F. C. 2013, ApJ, 778, 169

\bibitem[Batygin et al.(2015)]{Batygin2015}Batygin, K., Bodenheimer, P. H., \& Laughlin, G. P. (2015), arXiv preprint, arXiv:1511.09157.

\bibitem[Beaug\'e \& Nesvorn\'y(2012)]{Beauge2012}Beaug\'e, C., \& Nesvorn\'y, D. (2012), ApJ, 751(2), 119.

\bibitem[Becker \& Adams(2016)]{Becker2016}Becker, J. C., \& Adams, F. C. (2016), MNRAS, 455(3), 2980-2993.

\bibitem[Beuther et al.(2014)]{Beuther2014}Beuther, H., Klessen, R. S., Dullemond, C. P., \& Henning, T. K. (Eds.). (2014). Protostars and Planets VI. University of Arizona Press.

\bibitem[Boley et al.(2016)]{Boley2016}Boley, A. C., Contreras, A. G., \& Gladman, B. (2016), ApJL, 817(2), L17.

\bibitem[Bouvier(2013)]{Bouvier2013}Bouvier, J. (2013). EAS Publications Series, 62, 143.

\bibitem[Chambers(1999)]{Chambers1999}Chambers, J. E. (1999), ApJ, 304, 793.

\bibitem[Chandrasekar(1939)]{Chandrasekar1939} Chandrasekar, S. 1939, An Introduction to the Study of Stellar Structure
(Chicago: Univ. Chicago Press)

\bibitem[Cresswell et al.(2007)]{Cresswell2007}Cresswell, P., Dirksen, G., Kley, W., \& Nelson, R. P. (2007), A\&A, 473(1), 329.

\bibitem[Danby(1992)]{Danby1992}Danby, J. (1992). Fundamentals of celestial mechanics. Richmond: Willman-Bell,| c1992, 2nd ed., 1.

\bibitem[Dawson \& Chiang(2014)]{Dawson2014a}Dawson, R. I., \& Chiang, E. (2014), Science, 346, 212.

\bibitem[Dawson et al.(2014)]{Dawson2014b}Dawson, R. I., Murray-Clay, R. A., \& Johnson, J. A. (2014), ApJ, 798(2), 66.

\bibitem[Duch\^ene \& Kraus(2013)]{Duchene2013}Duch\^ene, G., \& Kraus, A. (2013), ARA\&A, 51, 269.

\bibitem[Fabrycky et al.(2014)]{Fabrycky2014}Fabrycky, D. C., Lissauer, J. J., Ragozzine, D., Rowe, J. F., Steffen, J. H., Agol, E., ... \& Ford, E. B. (2014), ApJ, 790(2), 146.

\bibitem[Fielding et al.(2015)]{Fielding2015}Fielding, D. B., McKee, C. F., Socrates, A., Cunningham, A. J., \& Klein, R. I. (2015), MNRAS, 450(3), 3306.

\bibitem[Ford \& Rasio(2008)]{Ford2008}
Ford, E. B., \& Rasio, F. A. 2008, ApJ, 686, 621  

\bibitem[Fragner \& Nelson(2010)]{Fragner2010}Fragner, M. M., \& Nelson, R. P. (2010), A\&A, 511, A77.

\bibitem[Gregory et al.(2012)]{Gregory2012}Gregory, S. G., Donati, J. F., Morin, J., Hussain, G. A. J., Mayne, N. J., Hillenbrand, L. A., \& Jardine, M. (2012), ApJ, 755(2), 97.

\bibitem[Huber et al.(2013)]{Huber2013}Huber, D., Carter, J. A., Barbieri, M., Miglio, A., Deck, K. M., Fabrycky, D. C., ... \& Winn, J. N. (2013), Science, 342(6156), 331-334.

\bibitem[Irwin et al.(2008)]{Irwin2008}Irwin, J., Hodgkin, S., Aigrain, S., Bouvier, J., Hebb, L., \& Moraux, E. (2008), MNRAS, 383, 1588.

\bibitem[Jensen \& Akeson(2014)]{Jensen2014}Jensen, E. L., \& Akeson, R. (2014), Nature, 511(7511), 567.

\bibitem[Johansen et al.(2012)]{Johansen2012}Johansen, A., Davies, M. B., Church, R. P., \& Holmelin, V. (2012), ApJ, 758, 39.

\bibitem[Kant(1755)]{Kant1755}Kant, I., 1755, General History of Nature and Theory of the Heavens (K\"onigsberg:Petersen)

\bibitem[Kley \& Nelson(2012)]{Kley2012}Kley, W., \& Nelson, R. P. (2012), ARA\&A, 50, 211.

\bibitem[Lai(2012)]{Lai2012}Lai, D. (2012). \textit{MNRAS}, 423(1), 486.

\bibitem[Lai(2014)]{Lai2014}Lai, D. (2014), MNRAS, 440, 3532.

\bibitem[Laplace(1796)]{Laplace1796}Laplace, P., S., 1796, Exposition dy syst\`eme du monde (Paris: Cerie-Social)

\bibitem[Lasker(1996)]{Laskar1996}Laskar, J. (1996), Cel. Mech \& Dyn. Ast., 64(1-2), 115.

\bibitem[Laughlin et al.(2002)]{Laughlin2002}Laughlin, G., Chambers, J., \& Fischer, D. (2002), ApJ, 579, 455.

\bibitem[Li \& Winn(2016)]{Li2016}Li, G., \& Winn, J. N. (2016), ApJ, 818(1), 5.

\bibitem[Lichtenberg \& Lieberman(1992)]{Lichtenberg1992}Lichtenberg, A. J., \& Lieberman, M. A. (1992). Regular and chaotic dynamics, volume 38 of Applied mathematical sciences.

\bibitem[Lithwick \& Wu(2012)]{Lithwick2012} 
Lithkwick, Y., \& Wu, Y. 2012, ApJ, 756, L11

\bibitem[Lissauer et al.(2011)]{Lissauer2011a}Lissauer, J. J., Fabrycky, D. C., Ford, E. B., Borucki, W. J., Fressin, F., Marcy, G. W., ... \& Batalha, N. M. (2011), Nature, 470, 53.

\bibitem[Lissauer et al.(2011)]{Lissauer2011b}Lissauer, J. J., Ragozzine, D., Fabrycky, D. C., Steffen, J. H., Ford, E. B., Jenkins, J. M., ... \& Batalha, N. M. (2011), ApJSS,197, 8.

\bibitem[Lissauer et al.(2013)]{Lissauer2013}Lissauer, J. J., Jontof-Hutter, D., Rowe, J. F., Fabrycky, D. C., Lopez, E. D., Agol, E., ... \& Howell, S. B. (2013), ApJ, 770, 131.

\bibitem[Mazeh et al.(2015)]{Mazeh2015}Mazeh, T., Perets, H. B., McQuillan, A., \& Goldstein, E. S. (2015), ApJ, 801(1), 3.

\bibitem[McQuillan et al.(2013)]{McQuillan2013}McQuillan, A., Mazeh, T., \& Aigrain, S. (2013), ApJL, 775(1), L11.

\bibitem[Morbidelli(2002)]{Morby2002}Morbidelli, A. (2002). Modern celestial mechanics: aspects of solar system dynamics (Vol. 1).

\bibitem[Morton \& Winn(2014)]{Morton2014}Morton, T. D., \& Winn, J. N. (2014), ApJ, 796, 47.

\bibitem[Murray \& Dermott(1999)]{Murray1999}Murray, C. D., \& Dermott, S. F. (1999). Solar system dynamics. Cambridge university press.

\bibitem[Naoz et al.(2011)]{Naoz2011}
Naoz, S., Farr, W. M., Lithwick, Y., Rasio, F. A., \& 
Teyssandier, J. 2011, Nature, 473, 187

\bibitem[Nelson et al.(2014)]{Nelson2014}Nelson, B. E., Ford, E. B., Wright, J. T., Fischer, D. A., von Braun, K., Howard, A. W., ... \& Dindar, S. (2014), MNRAS, 441, 442.

\bibitem[Nobili \& Roxburgh(1986)]{Nobili1986}Nobili, A., \& Roxburgh, I. W. (1986), In Relativity in Celestial Mechanics and Astrometry. High Precision Dynamical Theories and Observational Verifications (Vol. 114, pp. 105-110).

\bibitem[Petrovich(2015)]{Petrovich2015a}Petrovich, C. (2015), ApJ, 799, 27.

\bibitem[Petrovich(2015)]{Petrovich2015b}Petrovich, C. (2015), ApJ, 805, 75.

\bibitem[Shu et al.(1987)]{Shu1987} 
Shu, F. C., Adams, F. C., \& Lizano, S. 1987, ARA\&A, 25, 23

\bibitem[Siess et al.(2000)]{Siess2000}Siess, L., Dufour, E., \& Forestini, M. (2000), A\&A, 358, 593.

\bibitem[Sivia(1996)]{Sivia1996}Sivia, D. S. (1996). Data analysis: a Bayesian tutorial. Oxford university press.

\bibitem[Spalding \& Batygin(2014)]{Spalding2014a}Spalding, C., \& Batygin, K. (2014), ApJ, 790, 42.

\bibitem[Spalding et al.(2014)]{Spalding2014b}Spalding, C., Batygin, K., \& Adams, F. C. (2014), ApJL, 797, L29.

\bibitem[Spalding \& Batygin(2015)]{Spalding2015}Spalding, C., \& Batygin, K. (2015), ApJ, 811, 82.

\bibitem[Sterne(1939)]{Sterne1939}Sterne, T. E. (1939), MNRAS, 99, 451.

\bibitem[Storch et al.(2014)]{Storch2014}Storch, N. I., Anderson, K. R., \& Lai, D. (2014). \textit{Science}, 345(6202), 1317.

\bibitem[Tanaka \& Ward(2004)]{Tanaka2004}Tanaka, H., \& Ward, W. R. (2004), ApJ, 602(1), 388.

\bibitem[Veras \& Armitage(2004)]{Veras2004}Veras, D., \& Armitage, P. J. (2004), Icarus, 172, 349.

\bibitem[Winn et al.(2010)]{Winn2010}Winn, J. N., Fabrycky, D., Albrecht, S., \& Johnson, J. A. (2010), ApJL, 718, L145.

\bibitem[Winn \& Fabrycky(2015)]{Winn2015}Winn, J., \& Fabrycky, D. C. (2015), ARA\&A, 53.

\bibitem[Wu \& Murray(2003)]{Wu2003}Wu, Y., \& Murray, N. (2003), ApJ, 589(1), 605.
\end{thebibliography}
\end{document}